# Burst-Mode Ultrafast Laser Welding of Sapphire and Invar Alloy Across Large Interfacial Gaps up to 10 μm

**Yuxuan Li[1,2], Nan Li[1*], Yu Wang[1], Yitong Chen[1,3], Rong Su[1], Qingwei Zhang[1], Rongxian Wen[1,2], Guochang Jiang[1], Feng Chen[4], Shanglu Yang[1*]**

[1] Department of Advanced Optical and Microelectronic Equipment, Shanghai Institute of Optics and Fine Mechanics, Chinese Academy of Sciences, Shanghai, 201800, China
[2] Center of Materials Science and Optoelectronics Engineering, University of Chinese Academy of Sciences, Beijing, 100049, China
[3] College of Telecommunications and Information Engineering, Nanjing University of Posts and Telecommunications, Nanjing, 210003, China
[4] School of Fiber Engineering and Equipment Technology, Jiangnan University, Wuxi, 214122, China
*Corresponding author: yangshanglu@siom.ac.cnand nanli@siom.ac.cn

## Abstract

Achieving reliable joining between transparent materials and metals under non-optical-contact conditions remains challenging due to limited energy coupling and uncontrolled interfacial reaction across μm-scale gaps. Burst-mode ultrafast lasers provide a potential solution for large-gap welding through temporally distributed energy deposition. However, the underlying interaction mechanisms and achievable joining limits remain unclear. In this study, burst-mode ultrafast laser welding of sapphire to Invar alloy was investigated under controlled interfacial gaps from 3 to 10 μm. Cross-sectional microscopy, elemental mapping, white-light interferometry, and shear testing were employed to analyze joint morphology, elemental distribution, fracture behavior, and mechanical performance.After optimization of the processing parameters for burst-mode ultrafast laser welding, the interfacial morphological evolution and joint strength under different gap conditions were systematically investigated. At a 3 μm gap, cyclic thermal stresses induced by burst pulses generate transverse micro-crack networks in sapphire, accompanied by a reduction in joint strength with increasing sub-pulse numbers. Notably, at a 10 μm gap, where single-pulse welding fails, burst-mode ultrafast laser welding enables interfacial bridging with a maximum shear strength of 6.3 MPa, representing the highest level among published studies.These results indicate a gap-dependent evolution in burst-mode welding behavior governed by crack formation and energy accumulation. This study provides an important theoretical basis and practical guidance for achieving high-performance joining of dissimilar materials under large gap conditions.

## 1. Introduction

Sapphire ($\alpha$-$Al_2O_3$) possesses excellent mechanical, optical, and thermal properties, making it a preferred material for optical windows used in extreme environments. [1-2] Invar alloy (Fe–36Ni) exhibits an extremely low coefficient of thermal expansion and is widely used as high-precision structural material.[3] Reliably joining these two materials has significant potential for optoelectronic window packaging in aerospace, semiconductor, and medical applications.[4-5] However, their coefficients of thermal expansion differ by approximately 4.7 times, and other physical properties vary substantially, posing major challenges for high-quality bonding.[6-9] Conventional techniques have their own drawbacks: brazing introduces high heat input and residual stress;[10-11] adhesive bonding is susceptible to aging;[12] and mechanical joining requires complex assemblies.[13-14] These methods struggle to meet the demands of high-performance integration. Thus, there is a clear need for an advanced joining technique that enables efficient and clean joining while preserving structural and functional integrity.[15-16]

Ultrafast laser welding technology offers an innovative solution to this challenge.[17-19] In this technique, ultrafast laser pulses are focused at the joining interface, where nonlinear absorption enables highly localized energy deposition, interfacial melting, and plasma formation. This localized interaction promotes chemical bonding, metallurgical bonding, and mechanical interlocking, thereby enabling high-quality bonding with a minimal heat-affected zone and high joint strength.[20-23] In 2023, Pan *et al.* achieved femtosecond laser micro-welded joints of sapphire/Invar alloy without an interlayer under optical contact conditions.[24] Li *et al.* precisely controlled the assembly gap using white-light interferometry and systematically investigated the interfacial evolution during ultrafast laser welding of sapphire Invar alloy within the gap range of up to 2.4 μm. They found that as the gap increased, excessive melting and splashing of sapphire led to increased crack density, deterioration of joint integrity, and a decline in joining strength.[25]

Traditional ultrafast laser welding without burst-mode shows very limited tolerance for interfacial gaps.[26-30] The burst-mode pulse divides a single pulse into a series of sub-pulses. Studies indicate that this approach can significantly improve gap tolerance in ultrafast laser welding.[31-35] For example, Chen *et al.* used a femtosecond laser burst-mode to bond diamond and glass under a non- optical- contact condition of 3 μm, achieving a strength of 14.1 MPa. During the bonding process, molten glass jets filling the gap and forming laser- induced periodic surface structures (LIPSS) on the diamond, which enhances mechanical interlocking.[36] Jia *et al.* successfully joined sapphire and Invar alloy with a surface roughness of Sa = 2.128 μm using a femtosecond burst-mode. High- speed imaging revealed that molten metal actively fills the gap and undergoes reverse growth, exciting sapphire plasma. The mixture fills the interface, forming a uniform T- shaped weld seam, with a joint strength of 9–11.73 MPa.[37]

Current studies on burst-mode laser welding have mainly reported phenomenological observations, while systematic analysis of the underlying interaction mechanism remains limited. In particular, the coupling between the non-optical-contact gap and the burst pulse sequence, as well as its influence on joint formation and mechanical strength, is still not well understood. In this study, burst-mode picosecond laser welding of sapphire/Invar alloy was systematically investigated under controlled non-optical-contact gaps. First, at a 3 μm gap, the welding process window was optimized by adjusting the pulse energy and the number of sub-pulses, and representative parameters for stable joining were determined. Subsequently, based on the optimized parameters, the interfacial gap was increased from 3 to 10 μm to investigate the evolution of joint morphology, interfacial bonding characteristics, fracture behavior, and mechanical performance with increasing gap size. Furthermore, under the large 10 μm non-optical-contact gap, the achievable joint strength and failure characteristics of burst-mode welding were specifically evaluated to clarify the boundary of its large-gap joining capability. Finally, by correlating interfacial microstructure, fracture behavior, and mechanical performance, the joint formation mechanism was discussed, with particular emphasis on the gap-dependent transition from burst-induced crack damage at small gaps to burst-enhanced interfacial bridging at large gaps. This work provides a mechanistic basis for process optimization and gap-tolerant burst-mode welding of transparent/metal dissimilar materials under micron-scale non-optical-contact conditions.

## 2. Results and Discussion

### 2.1 Optimization of Burst-Mode Ultrafast Laser Welding Parameters under a 3 μm Gap Condition

A 3 μm interfacial gap is close to the joining limit of ultrafast laser welding without burst-mode pulses.[25] Thus, this gap condition was used as the baseline for optimizing the burst-mode welding process. The effects of sub-pulse number and pulse energy on joint formation, interfacial morphology, crack evolution, fracture behavior, and ultimate shear strength were systematically examined to determine suitable processing parameters for large-gap welding.

#### 2.1.1 Effects of Sub-Pulse Number and Pulse Energy on Joint Formation

The evolution of joint morphology at the sapphire/Invar interface with an initial gap of 3 μm is shown in Figs. 1, 2. The process optimization was conducted by systematically investigating the synergistic effects of pulse energy and number of sub-pulses in burst-mode welding. Specifically, the pulse energy was varied from 55 to 130 μJ, while the number of sub-pulses, N, was adjusted under selected pulse-energy conditions to evaluate its influence on interfacial morphology and joint formation.

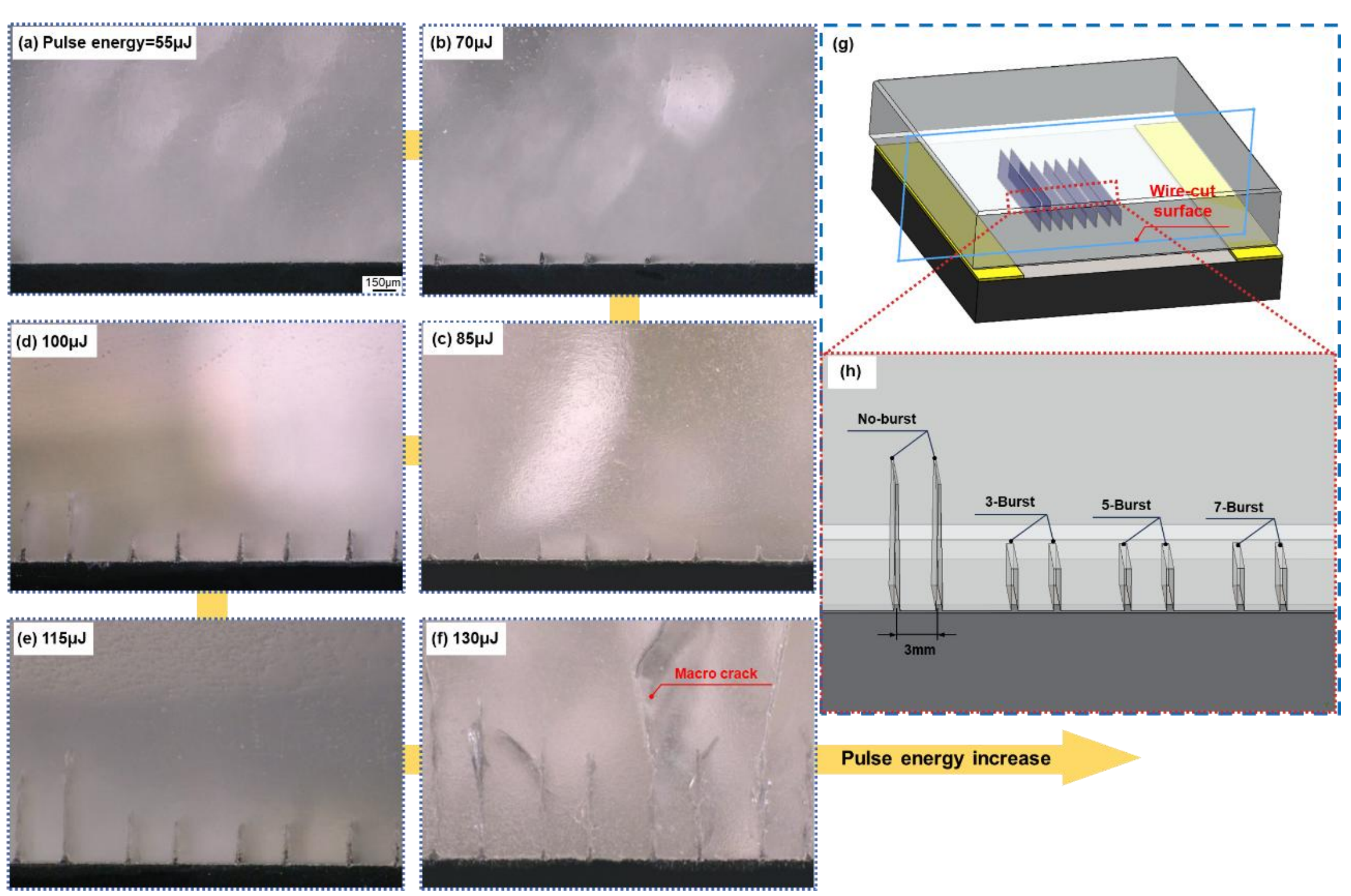

**Fig 1. Effects of sub-pulse number and pulse energy on joint formation under a 3 μm interfacial gap.** **a-f,** Cross-sectional optical micrographs of sapphire/Invar joints produced under different pulse energies and sub-pulse numbers. **g, h,** Schematics showing the wire-cutting positions and corresponding cross-sectional observation planes of the joints.

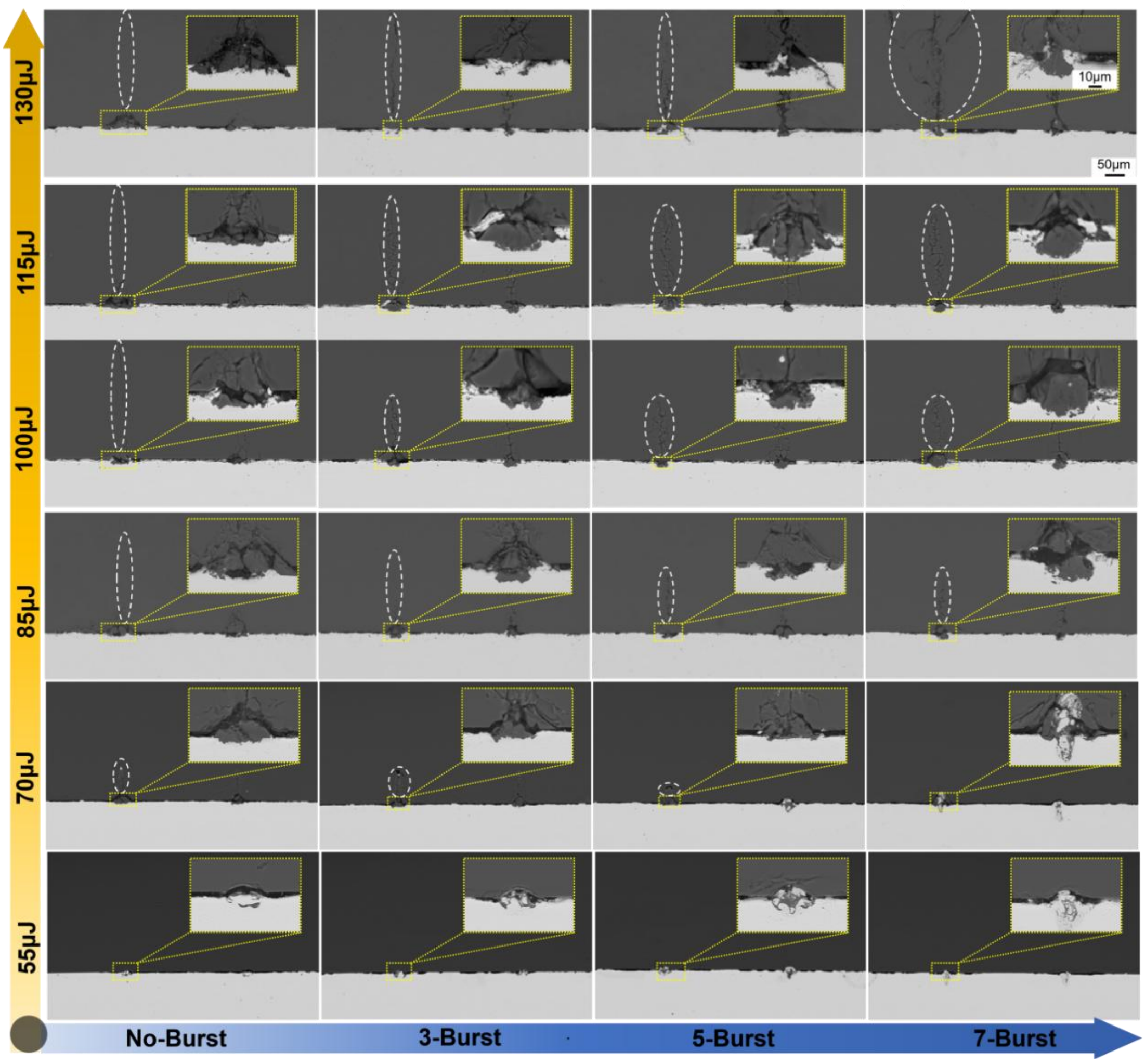

**Fig 2. Cross-sectional morphology of Burst-mode laser welds under different pulse energies and sub-pulse numbers.**

As shown in Fig. 2, when the pulse energy was below 70 μJ, only limited surface modification or isolated fusion points were observed, which were insufficient to establish continuous bonding. Compared with the no-burst mode, burst-mode irradiation generated more fusion points, indicating more effective interfacial energy utilization. At 85-115 μJ, continuous joint formation became evident. The no-burst mode produced a relatively wide and shallow fusion zone, whereas burst-mode irradiation formed a more continuous and compact bonding layer with a narrower and deeper morphology. This result indicates that temporally distributed sub-pulses improve molten-pool stability and interfacial reaction continuity. At 130 μJ, excessive energy input caused severe damage in all modes, and macroscopic cracking became evident under high-burst-number conditions.

At fixed pulse energies of 100 or 115 μJ, increasing N from 1 to 7 gradually transformed the joint morphology from localized pulse impact to cumulative melting and solidification. At N = 3 and 5, the bonding layer became more continuous and thicker, indicating that moderate thermal accumulation promotes interfacial fusion.

The SEM and EDS results in Fig. 3 further confirm that this morphological evolution is accompanied by a transition in bonding behavior. In the no-burst mode, the interface remained relatively sharp, suggesting limited elemental interdiffusion. In the burst-mode, the interface gradually evolved into a blurred transition layer. Al and O from sapphire diffused into the Invar side, while Fe and Ni penetrated into the sapphire-side reaction region, forming a micrometer-scale intermixing zone.

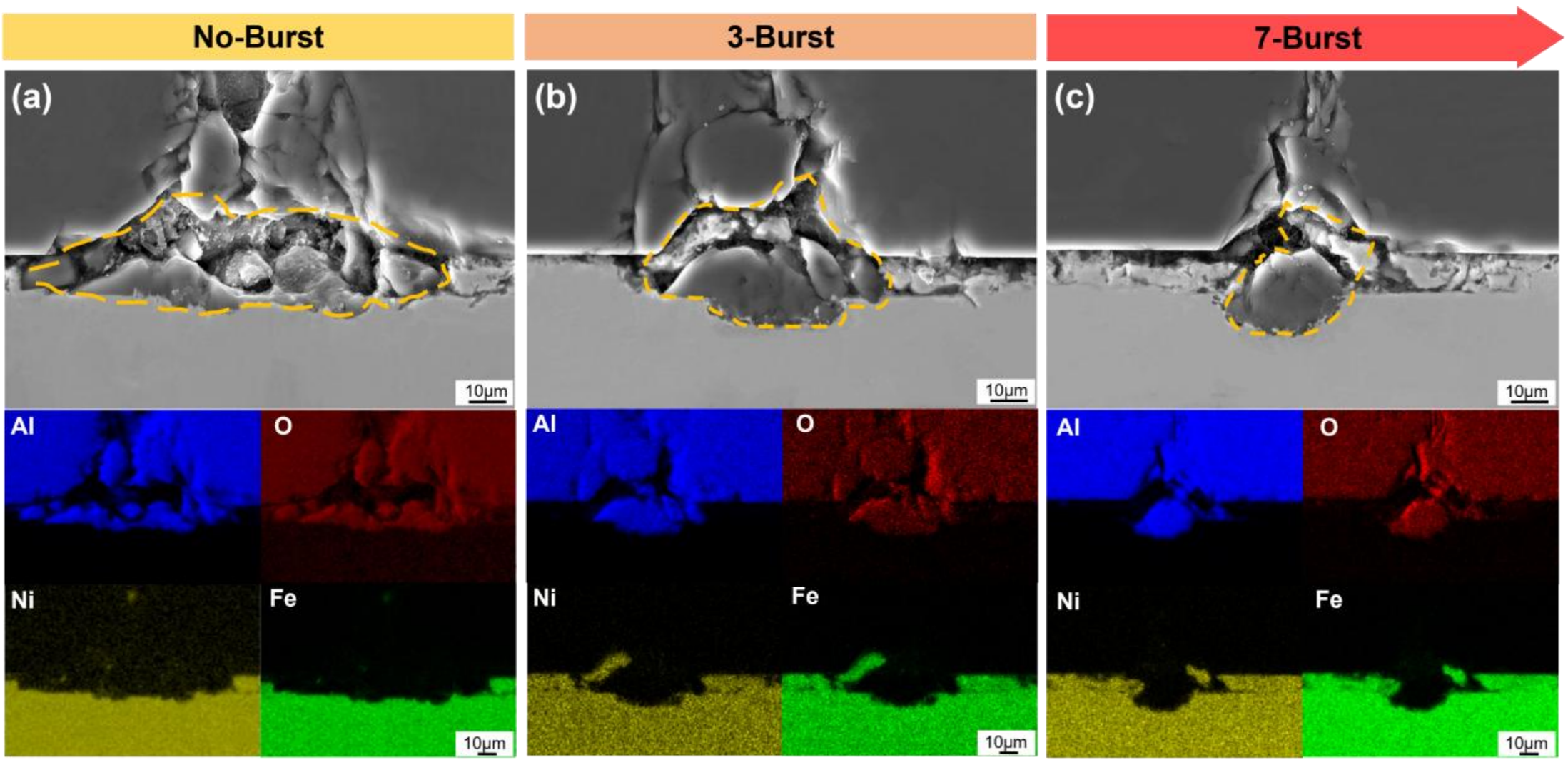

**Fig 3. Cross-sectional morphology and elemental distributions of sapphire/Invar joints produced with different sub-pulse numbers under a 3 μm interfacial gap. a,** No-Burst. **b,** 3-Burst. **c,** 7-Burst.

This transition indicates that burst-mode pulses modify interfacial bonding through multi-pulse thermal accumulation. Repeated heating-cooling cycles enhance melt flow and interfacial mixing, while sustained local heat input prolongs molten-pool lifetime and promotes elemental interdiffusion. As a result, burst-mode welding produces a denser and more continuous metallurgical bond.

### 2.1.2 Optimization Based on Interfacial Morphology and Shear Strength

Although burst-mode welding improves interfacial continuity at a 3 μm gap, the mechanical response exhibits the opposite trend. As shown in Fig. 4, the average shear strength decreases monotonically with increasing sub-pulse number. At 115 μJ, the joint strength drops from 32.2 MPa in the no-burst mode to 6.8 MPa in the 7-burst mode.

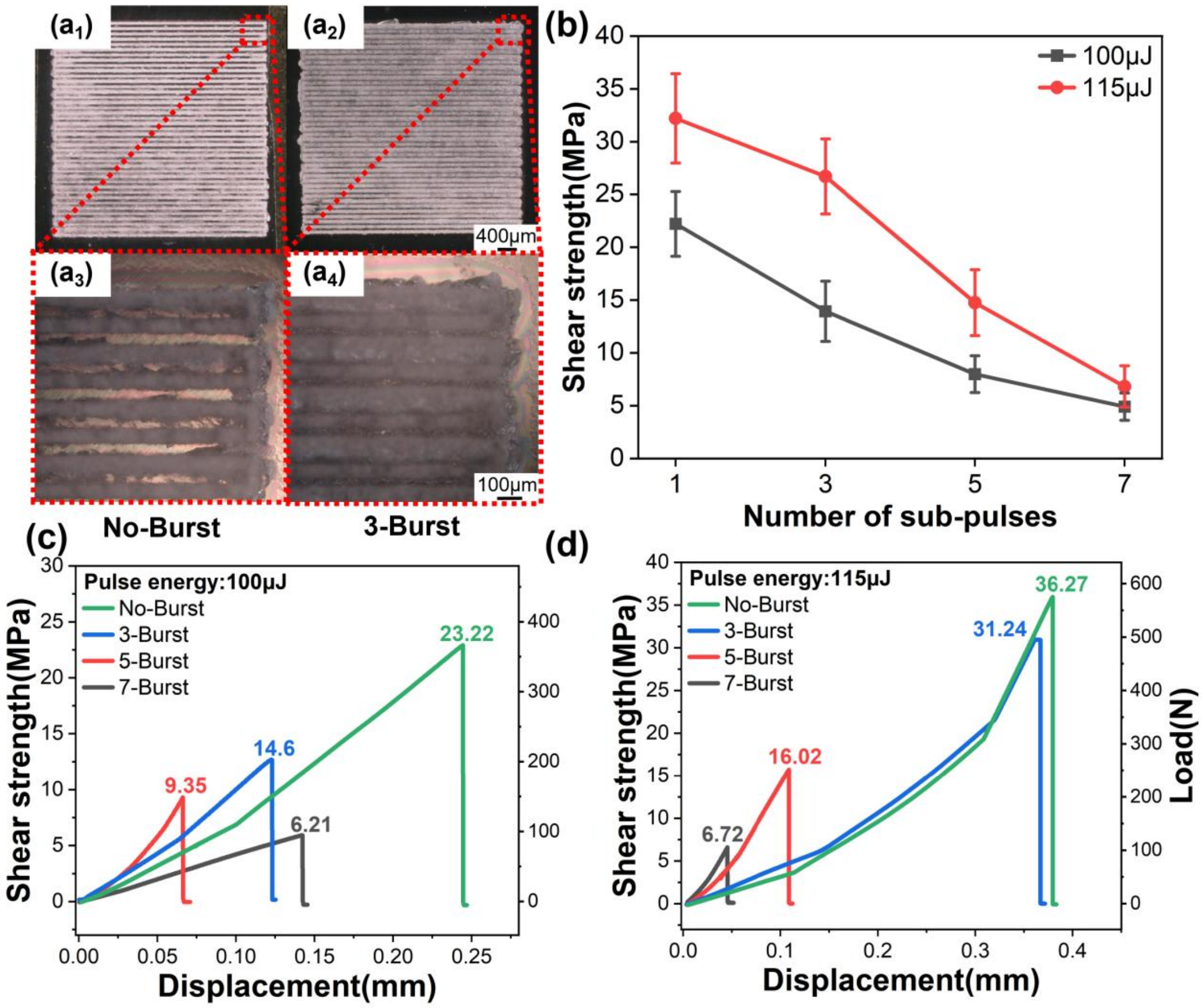

**Fig 4**. **Sub-pulse-number-dependent mechanical degradation of sapphire/Invar joints under a 3 μm gap condition. $a_1$-$a_4$**, Top view of optical micrographs at a pulse energy of 115 μJ. **b**, Shear strength as a function of pulse energy and number of sub-pulses. **c**, Strength displacement curve at 100 μJ. **d**, Strength displacement curve at 115 μJ.

This strength degradation is governed by the coupled evolution of subsurface cracking and fracture morphology, as jointly revealed in Fig. 5, 6. In the no-burst mode, only isolated microcracks are observed in the sapphire modified zone, and the fracture surface remains relatively flat, indicating localized brittle failure without an interconnected damage network. In contrast, burst-mode welding progressively forms a dense crack network composed of longitudinal primary cracks and transverse branches. With increasing sub-pulse number, these cracks propagate, deflect, and interconnect, ultimately producing a quasi-three-dimensional crack network within sapphire. This pre-existing damage network provides preferential fracture paths during shear loading and reduces the effective load-bearing area.

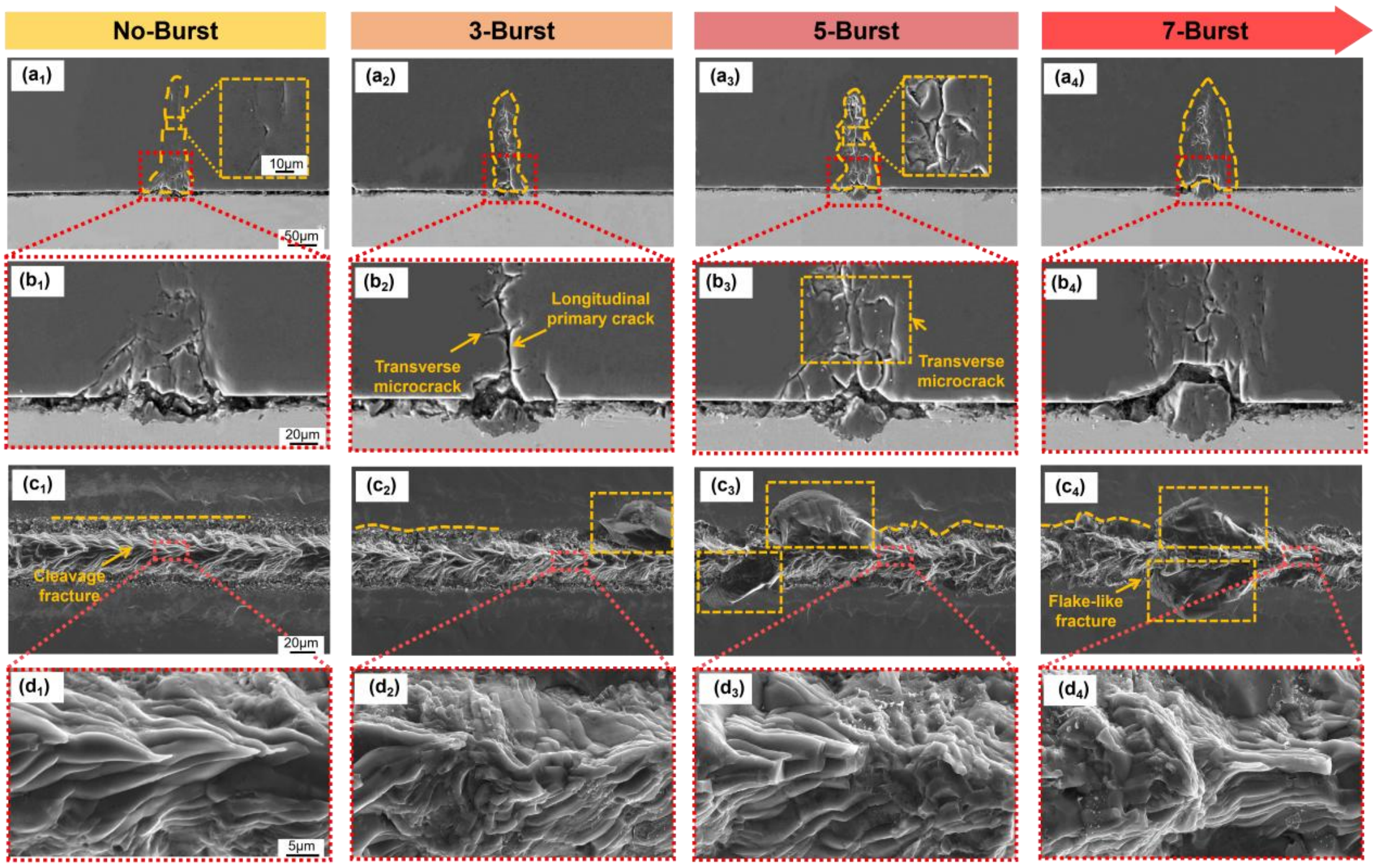

**Fig 5. Welded joints produced under different numbers of sub-pulses with a 3 μm interfacial gap. $a_1$-$a_4$, $b_1$-$b_4$,** Cross-sectional morphology of the joints. **$c_1$-$c_4$, $d_1$-$d_4$,** Fracture surfaces of the joints.

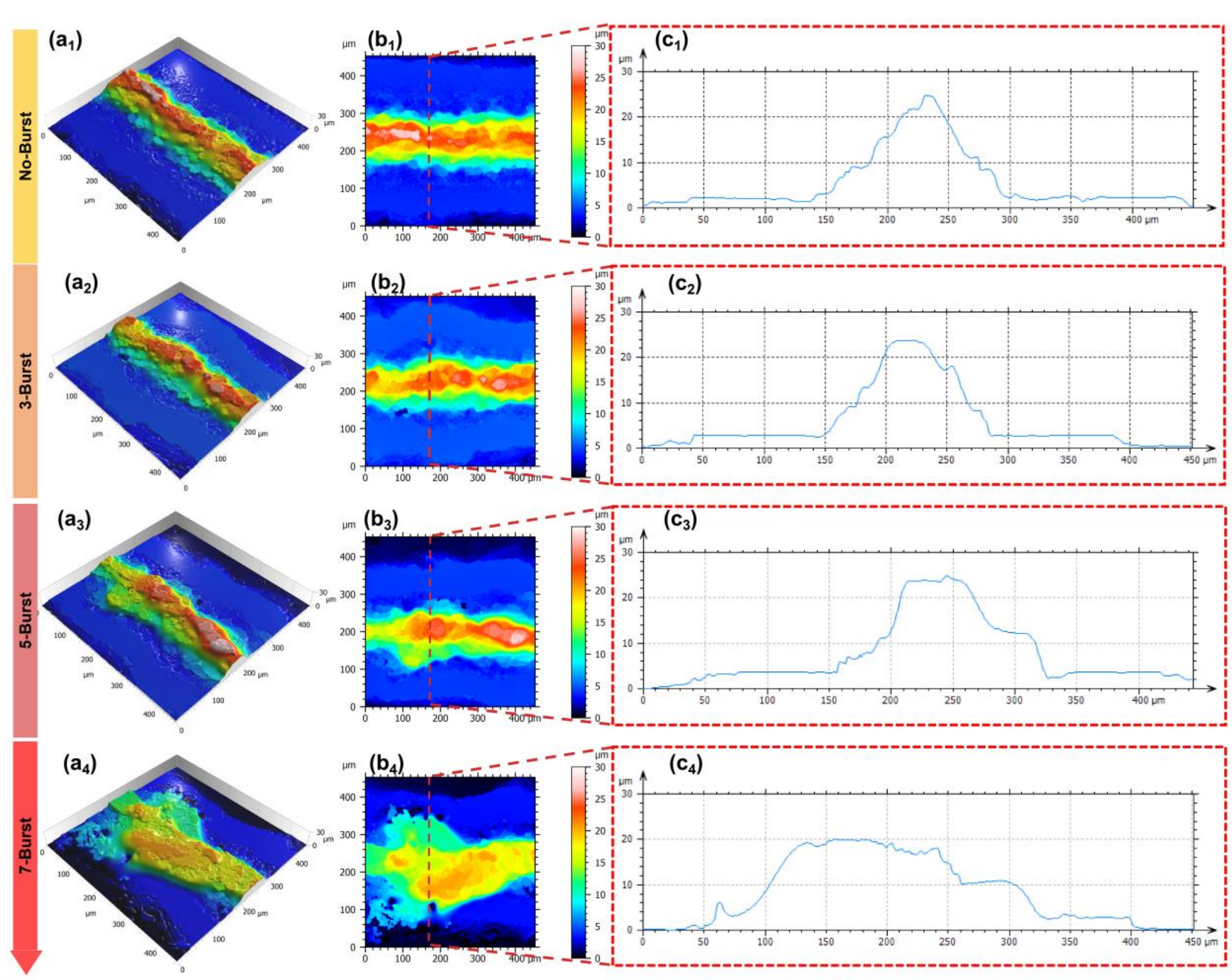

**Fig 6. White-light interferometric measurements of fracture surfaces obtained with different numbers of sub-pulses. $a_1$-$a_4$,** Representative three-dimensional fracture morphologies. **$b_1$-$b_4$,** Two-dimensional height contour maps of the corresponding fracture surfaces. **$c_1$-$c_4$,** Cross-sectional profiles extracted from the corresponding height maps.

In contrast, burst-mode progressively generates to a dense and interconnected crack network composed of longitudinal primary cracks and transverse branches. With increasing N, these cracks propagate, deflect, and interlink, ultimately forming a quasi-3D crack network within sapphire. This network directly governs the fracture path.This crack-network formation is consistent with the fracture surface evolution captured by white-light interferometry. In the no-burst mode, the fracture surface exhibits a sharp and localized peak, reflecting concentrated failure in a limited region. In contrast, burst-mode transforms the fracture topography into a broader, flatter multi-peak plateau with increasing roughness and lateral spreading. In the 7-burst mode, the surface becomes highly fragmented and discontinuous, consistent with crack-network-controlled-delamination rather than single-path fracture.

When the pulse width exceeds 10 ps, sapphire damage is governed primarily by energy deposition and thermal diffusion.[41-42] This strength degradation can be explained using a simplified fracture-mechanics framework.[43] Under burst-mode welding, repeated heating-cooling cycles impose cyclic thermal stress on the laser-modified sapphire region. The corresponding stress intensity range can be described as:

$$\Delta K = Y \cdot \Delta\sigma \cdot \sqrt{\pi a} \quad (1)$$

where Y is the geometry factor, $\Delta\sigma_{th}$ is the cyclic thermal stress amplitude induced by successive sub-pulses, and a is the characteristic crack length. Once ΔK exceeds the crack-growth threshold $\Delta K_{th}$, pre-existing or newly initiated cracks can be repeatedly reactivated during subsequent sub-pulse welding. As the crack length a increases, ΔK further increases, promoting crack extension, branching, and interconnection. This positive feedback process explains the transition from isolated microcracks in the no-burst mode to an interconnected crack network under high-burst-number welding.

This evolution is consistent with Figs. 5 and 6, where fracture morphology transitions from flat localized failure to increasingly rough and tortuous surfaces. At higher burst numbers, lamellar fracture with extensive flake-like sapphire fragments dominates, indicating crack-network-controlled failure rather than intrinsic material strength. Consequently, although interfacial continuity improves, increasing crack density reduces the effective load-bearing area, resulting in monotonic shear strength degradation.

**2.2 Gap-Dependent Evolution of Welded Joints**

When the interfacial gap exceeds 3 μm, energy attenuation caused by plasma shielding and air-gap propagation loss becomes increasingly severe, resulting in progressive failure of interfacial bonding in the no-burst mode. In contrast, burst-mode welding mitigates this limitation through temporally distributed energy deposition and cumulative sub-pulse heating, thereby improving gap tolerance and enabling effective joining at larger interfacial gaps.

### 2.2.1 Evolution of Interfacial Morphology Across 3-10 μm Gaps

A gap-dependent reversal in mechanical performance emerged with increasing interfacial gap. As shown in Fig. 7, at 6 μm the 3-burst mode began to outperform the no-burst mode in shear strength. This contrast becomes most evident at 10 μm, where the no-burst mode fails completely, whereas the 3-burst mode still maintains effective joining.

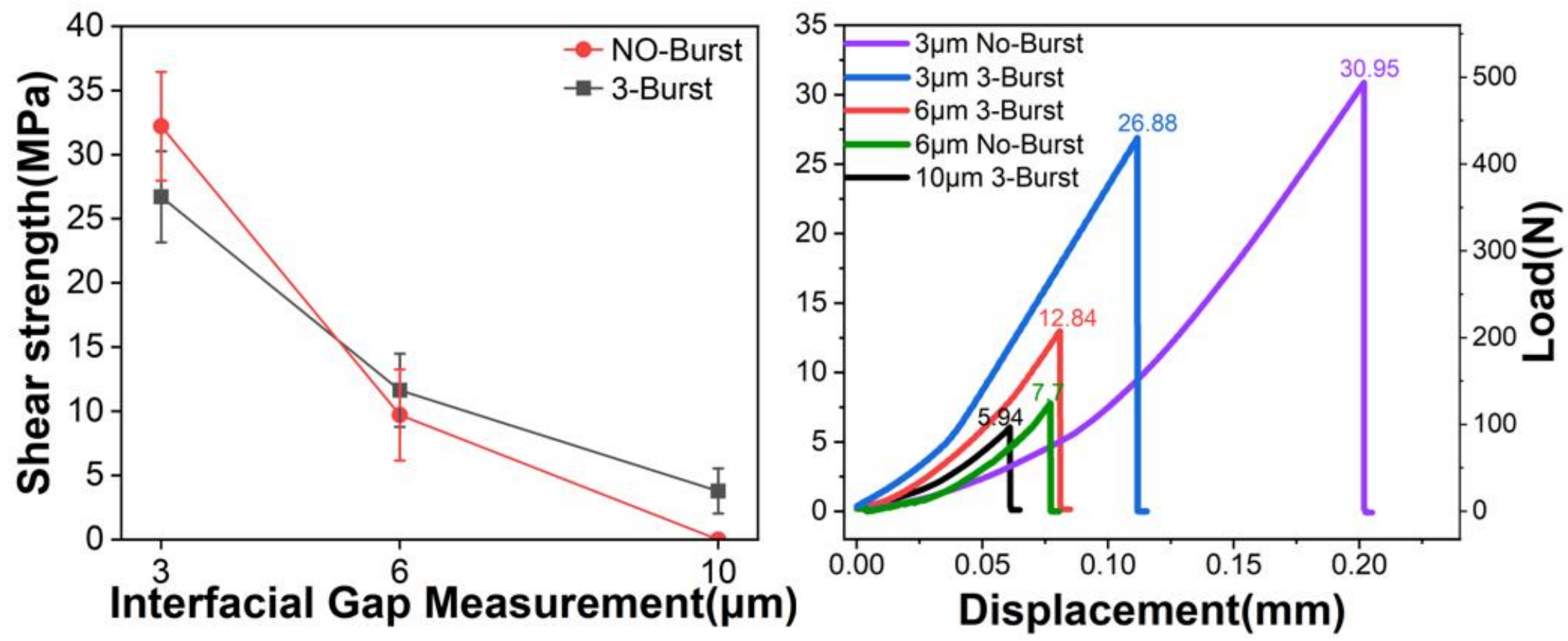

**Fig 7. Shear strength and strength-displacement curves of sapphire/Invar joints produced in the no-burst and 3-burst modes across 3-10 μm interfacial gaps.**

As shown in Fig. 8, at a 6 μm gap, the no-burst mode exhibited a discontinuous joining region with pores and lack-of-fusion regions, indicating insufficient energy delivery across the air gap. EDS maps revealed a discontinuous elemental distribution across the interface, confirming incomplete interfacial reaction. In contrast, burst-mode welding produced a more continuous bonding layer and a broader elemental interdiffusion zone, indicating that thermal accumulation effectively compensated for gap-induced energy attenuation and restored interfacial reaction continuity.

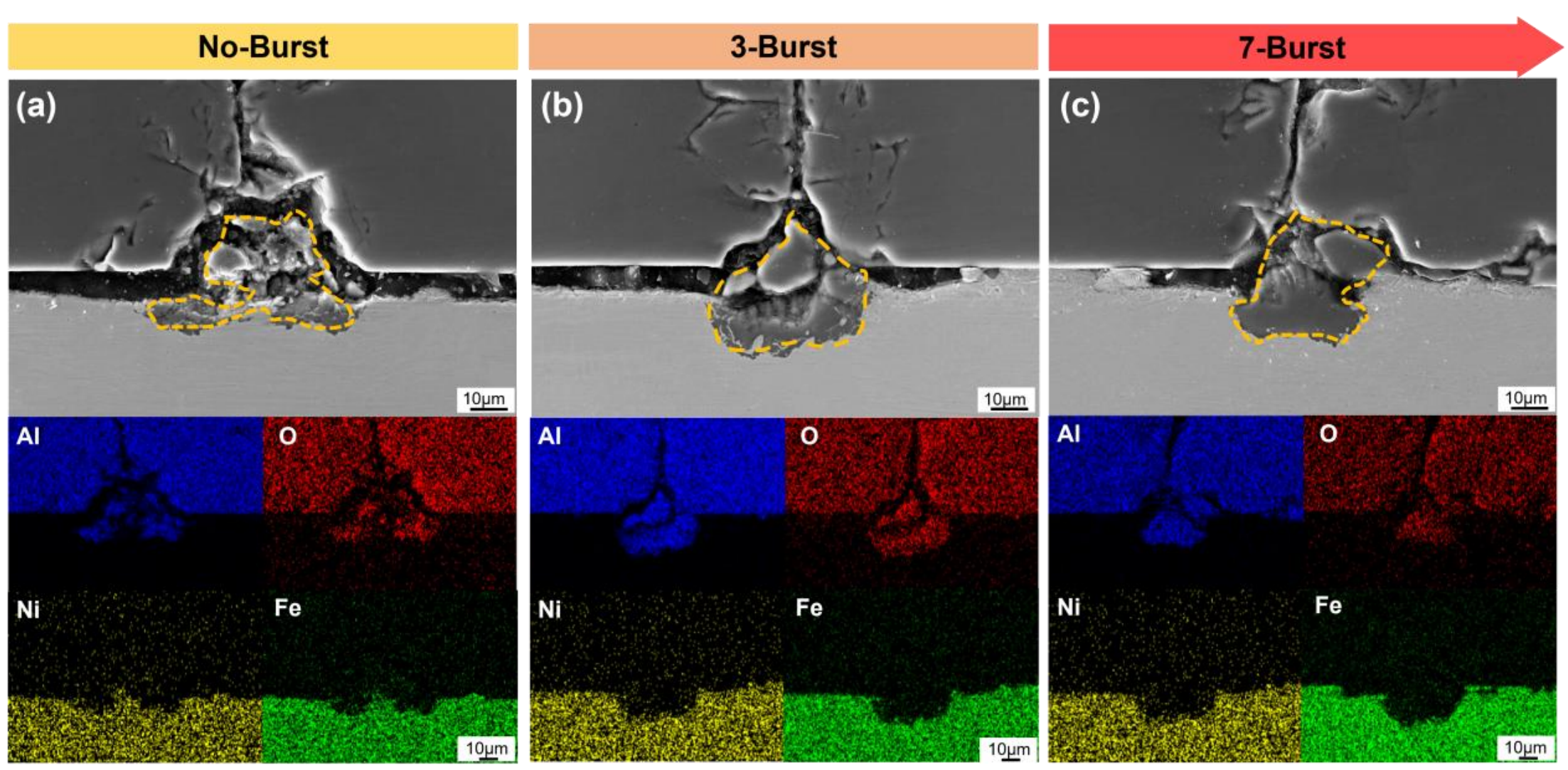

**Fig 8. Cross-sectional morphology and elemental distributions of sapphire/Invar joints produced with different sub-pulse numbers under a 6 μm interfacial gap. a,** No-Burst. **b,** 3-Burst. **c,** 7-Burst.

This transition was further confirmed by the cross-sectional morphology and elemental distribution under burst-mode welding. The burst-mode joints exhibited a continuous and dense bonding layer, together with a broad elemental interdiffusion zone. Al and O from sapphire diffused into the Invar side, while Ni and Fe penetrated into the sapphire-side reaction region, indicating sustained interfacial mixing across the 6 μm gap. Under this condition, the gap-tolerance advantage provided by burst-induced thermal accumulation outweighed the crack-damage effect in sapphire. As a result, the joint strength surpassed that of the no-burst mode, highlighting the critical role of burst-mode welding in large-gap joining.

As shown in Fig. 9, this difference became more pronounced at a 10 μm gap. In the no-burst mode, SEM revealed a distinct unbridged gap between sapphire and Invar, with only localized sapphire melting and no continuous fusion layer. The corresponding EDS maps showed a sharp elemental boundary, confirming the failure of effective joining. By contrast, burst-mode welding formed a continuous triangular transition zone bridging the two materials across the 10 μm gap. This reaction layer provided direct evidence that repeated sub-pulse heating enabled progressive molten-pool expansion and sustained interfacial mixing, thereby allowing the gap to be gradually filled and a continuous interfacial bond to form.

These results confirm that the advantage of burst-mode welding becomes increasingly pronounced with increasing interfacial gap, particularly near the 10 μm joining limit, where conventional no-burst welding can no longer achieve stable gap bridging.

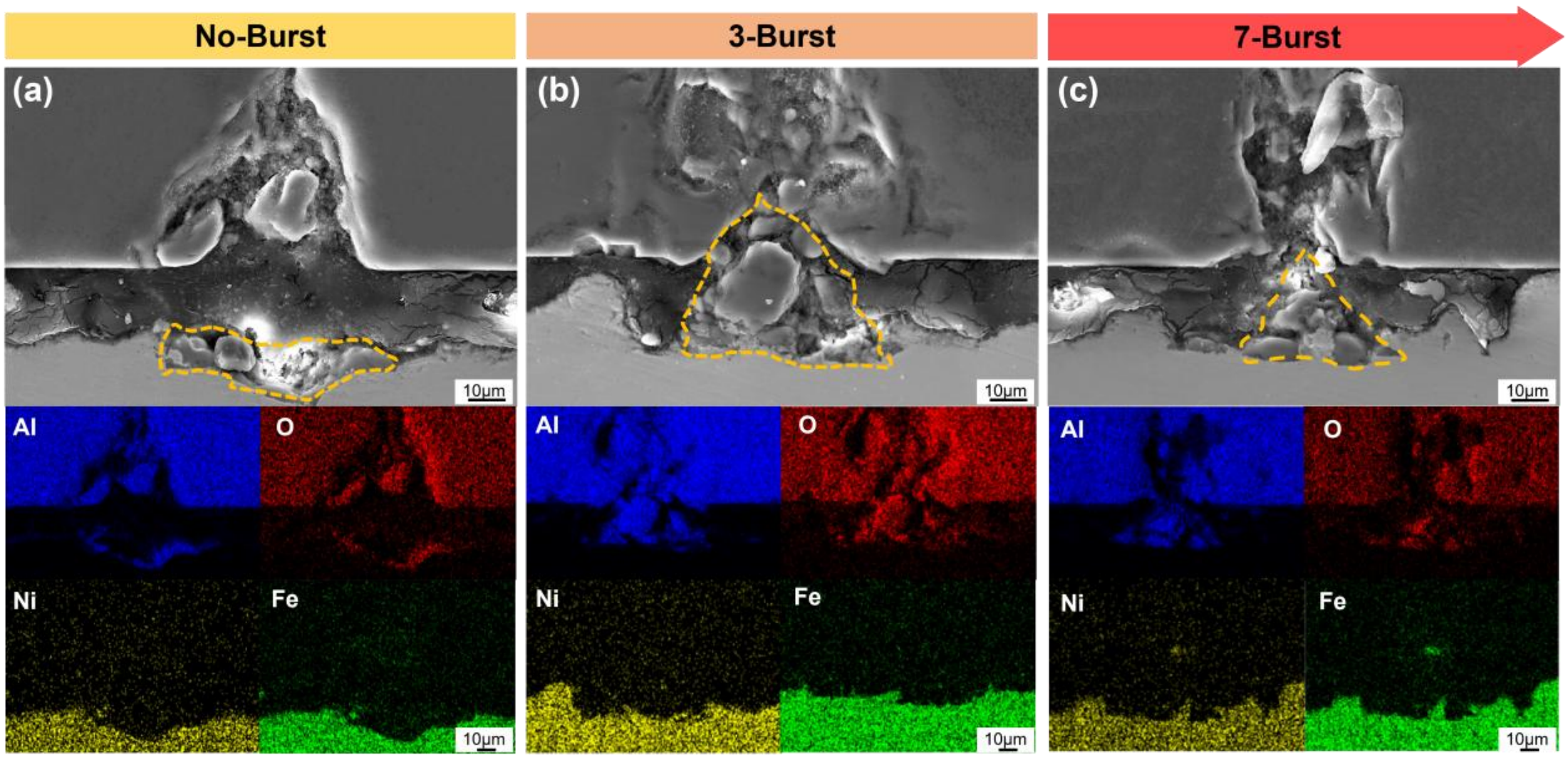

**Fig 9. Cross-sectional morphology and elemental distributions of sapphire/Invar joints produced with different sub-pulse numbers under a 10 μm interfacial gap. a,** No-Burst. **b,** 3-Burst. **c,** 7-Burst.

### 2.2.2 Fractural Analysis Under Varying Gap Conditions

To investigate the differences in joining behavior between the no-burst and 3-burst modes within the 6-10 μm gap range, welding was performed on gradient gap samples as shown in Fig. 10a-c.

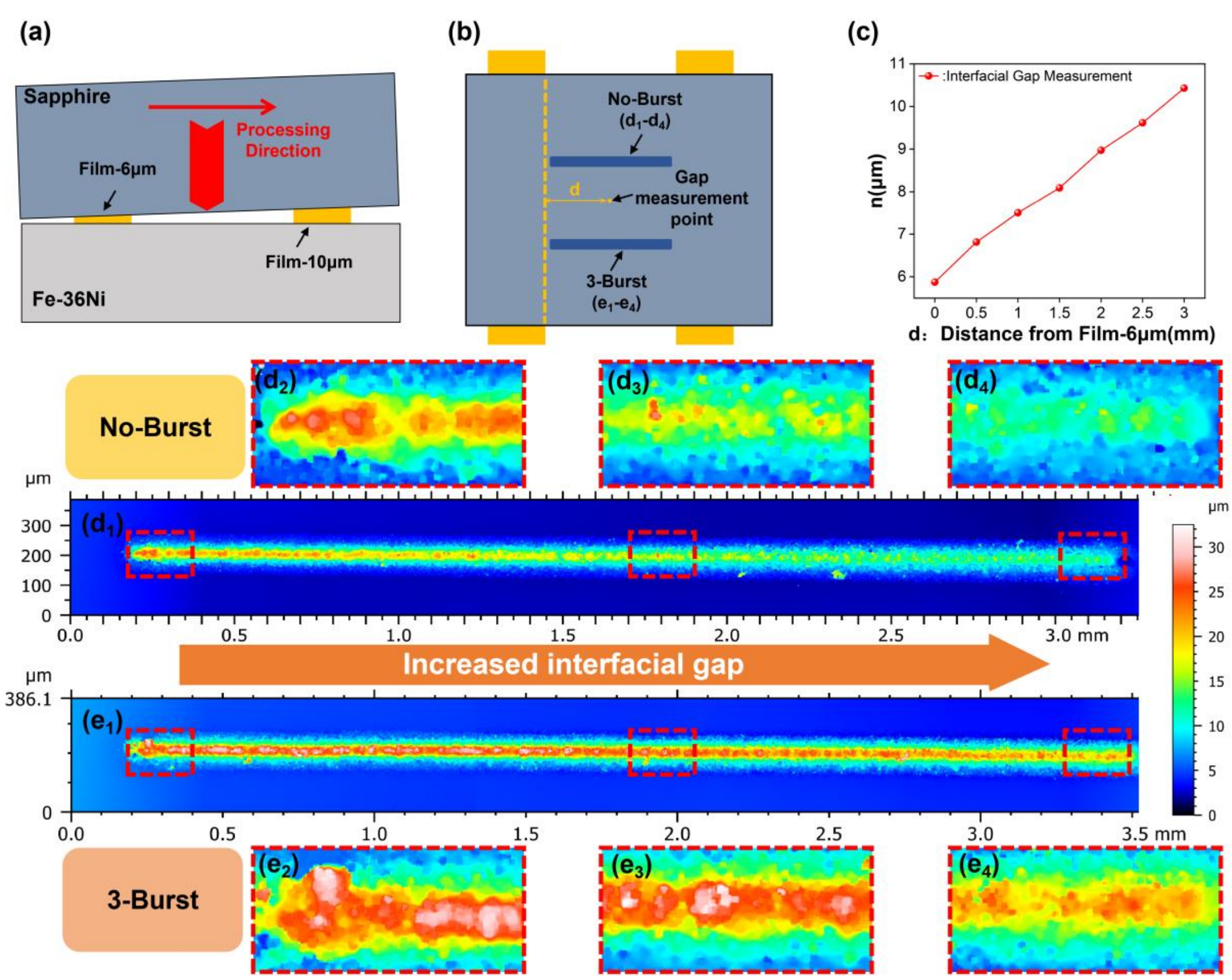

**Fig 10. Gap-dependent evolution fracture of topography under a continuously increasing interfacial gap. a-b,** Schematics of the interfacial gap regulation strategy. **c,** Schematic of the continuously increasing interfacial gap. **$d_1$-$d_4$, $e_1$-$e_4$,** White-light-interferometry height maps of the fracture surfaces after shear testing under different gap conditions.

In the no-burst mode, the fracture morphology evolved rapidly with increasing gap. A continuous bonded region was observed only near the small-gap end, whereas the bonded area quickly became discontinuous as the gap increased. At larger gaps, the fracture surface was dominated by sparse residual islands and shallow resolidified deposits, indicating progressive loss of effective interfacial bonding and a transition from continuous metallurgical bridging to localized melt attachment.

In contrast, the 3-burst mode maintained a substantially more continuous fracture morphology over the entire gradient range. Although the bonded region gradually became thinner as the gap increased, the central weld seam remained laterally continuous and retained measurable bonding thickness even near 10 μm. No abrupt interruption or island-like separation was observed, indicating that effective bonding was preserved over a substantially wider gap range.

These results further confirm that burst-mode welding compensated for gap-induced energy loss through cumulative sub-pulse heating and prolonged molten-pool lifetime, thereby extending the joining limit and preserving metallurgical continuity under large-gap conditions.

### 2.2.3 Welding Capability under 10 μm Interfacial Gaps

To evaluate the joining limit under the extreme 10 μm gap condition, fracture surfaces were systematically characterized after welding. In the no-burst mode, the fracture surface consisted mainly of isolated protrusions, pores, and resolidified sapphire droplets as shown in Fig. 11$a_1$, indicating only localized melting without effective metallurgical bridging.

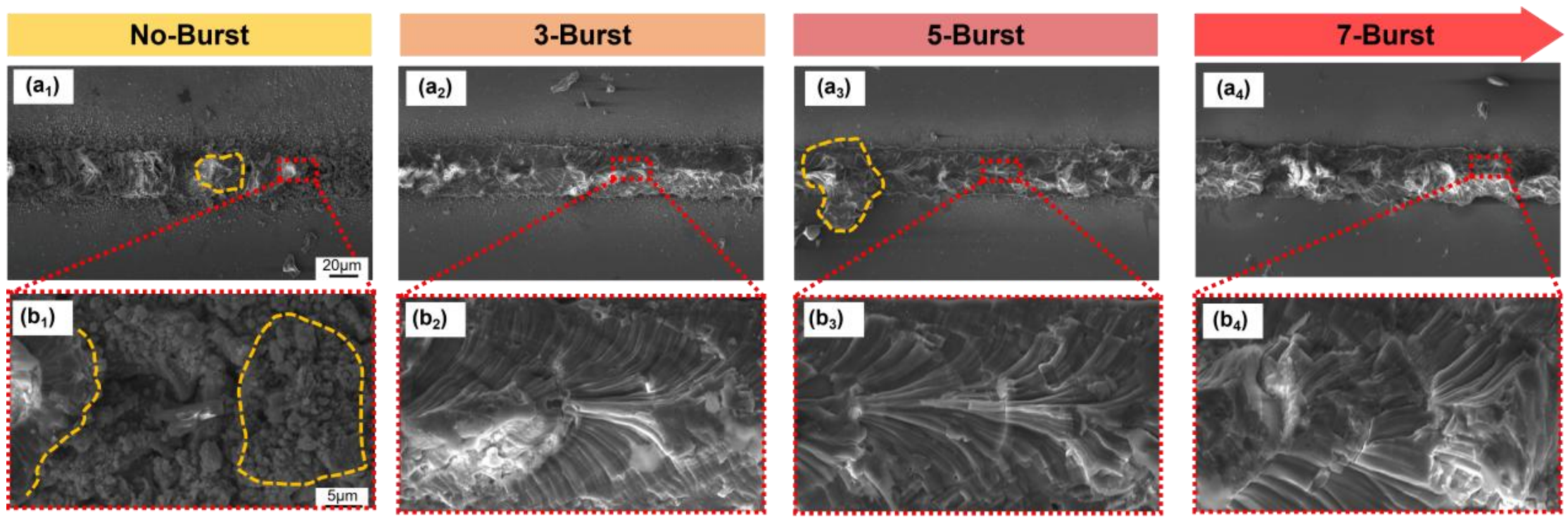

**Fig 11. Fracture-surface morphology of sapphire/Invar joints produced with different sub-pulse numbers under a 10 μm interfacial gap.**

In contrast, the burst-mode produced a continuous layered fracture morphology( Figs. 11$a_2$-$a_4$ and 11$b_2$-$b_4$), similar to that observed in successfully bonded joints at smaller gaps. This result indicated that repeated sub-pulse heating enabled sustained melting, intermixing, and layer-by-layer solidification, thereby establishing continuous bonding across the 10 μm gap. However, at higher numbers of sub-pulses, local sapphire spallation and irregular fracture features became evident, suggesting that excessive thermal cycling again introduced crack-assisted failure.

Further optimization under the 3-burst mode at a 10 μm gap showed that the effective joining window became significantly narrower than that at 3 μm. A higher pulse energy was required to compensate for transmission loss and provide sufficient molten material for gap filling. Accordingly, the optimal pulse energy shifted from 115 μJ at 3 μm to 130 μJ at 10 μm. Under this optimized condition, the weld seam exhibited continuous morphology and fracture features associated with effective interfacial bonding, as shown in Fig. 12. This result confirms that reliable burst-mode welding can still be achieved at the 10 μm large-gap limit.

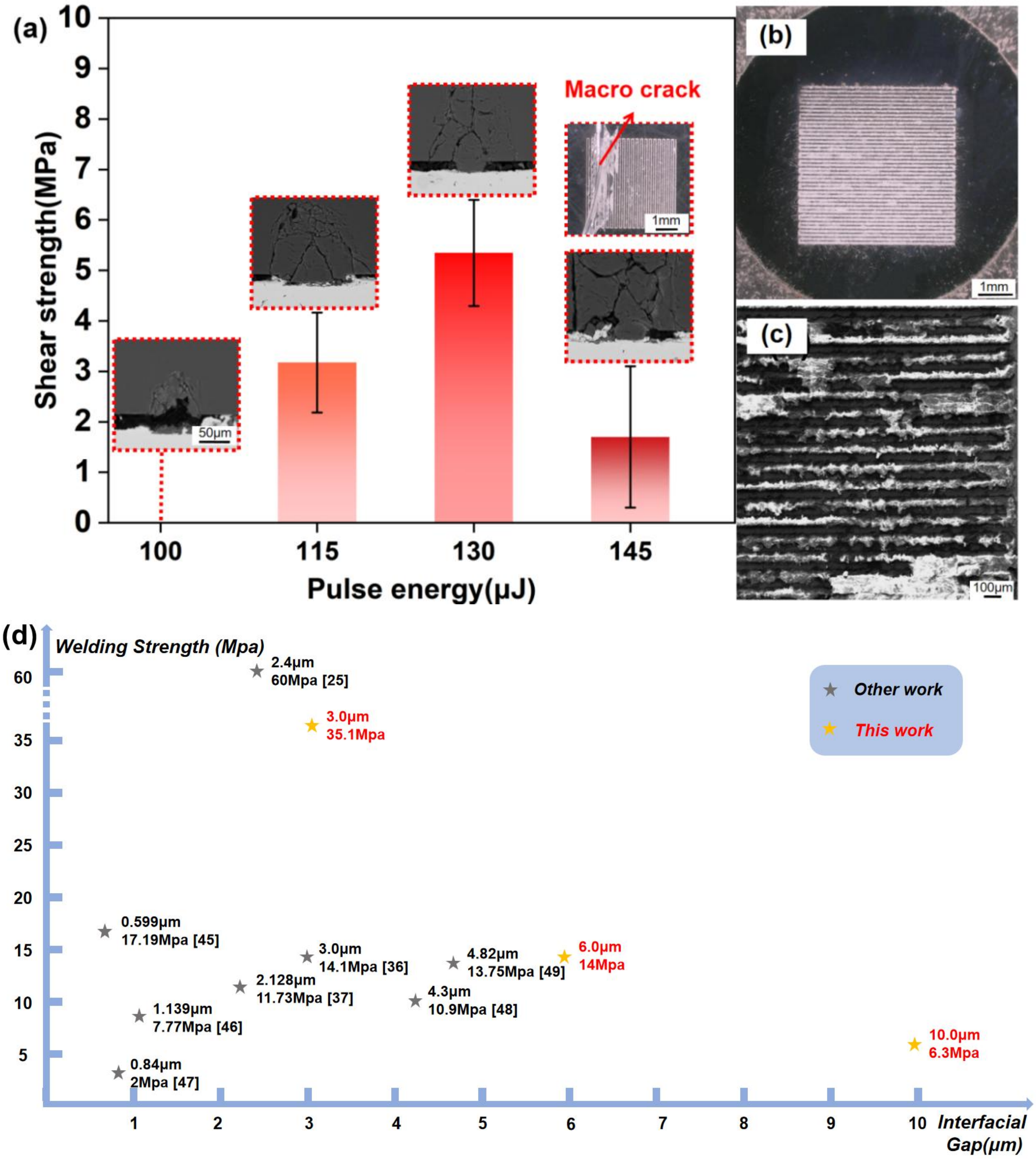

**Fig 12**. **Optimization of 3-burst welding at the 10 μm interfacial gap. a,** Shear strength of joints welded in the 3-Burst with a 10 μm gap, at different pulse energies. **b,** Optical micrograph

of the weld seam obtained at the optimized parameter of 130 μJ (3-Burst mode, 10 μm gap). **c,** Fracture surface micrograph of the corresponding joint. **d,** Comparison of ultrafast laser welding performance under different interfacial gap conditions.

## 2.3 Mechanism of Large-Gap Burst-Mode Welding

This study reveals that burst-mode ultrafast welding of sapphire and Invar is governed by a gap-dependent competition between two intrinsically coupled effects, as summarized in Fig. 13: (i) burst-induced crack damage in sapphire and (ii) burst-enhanced interfacial bridging that improves gap tolerance. These two mechanisms coexist during welding, but their relative dominance shifts systematically with increasing interfacial gap, thereby determining the final joint morphology and mechanical performance.

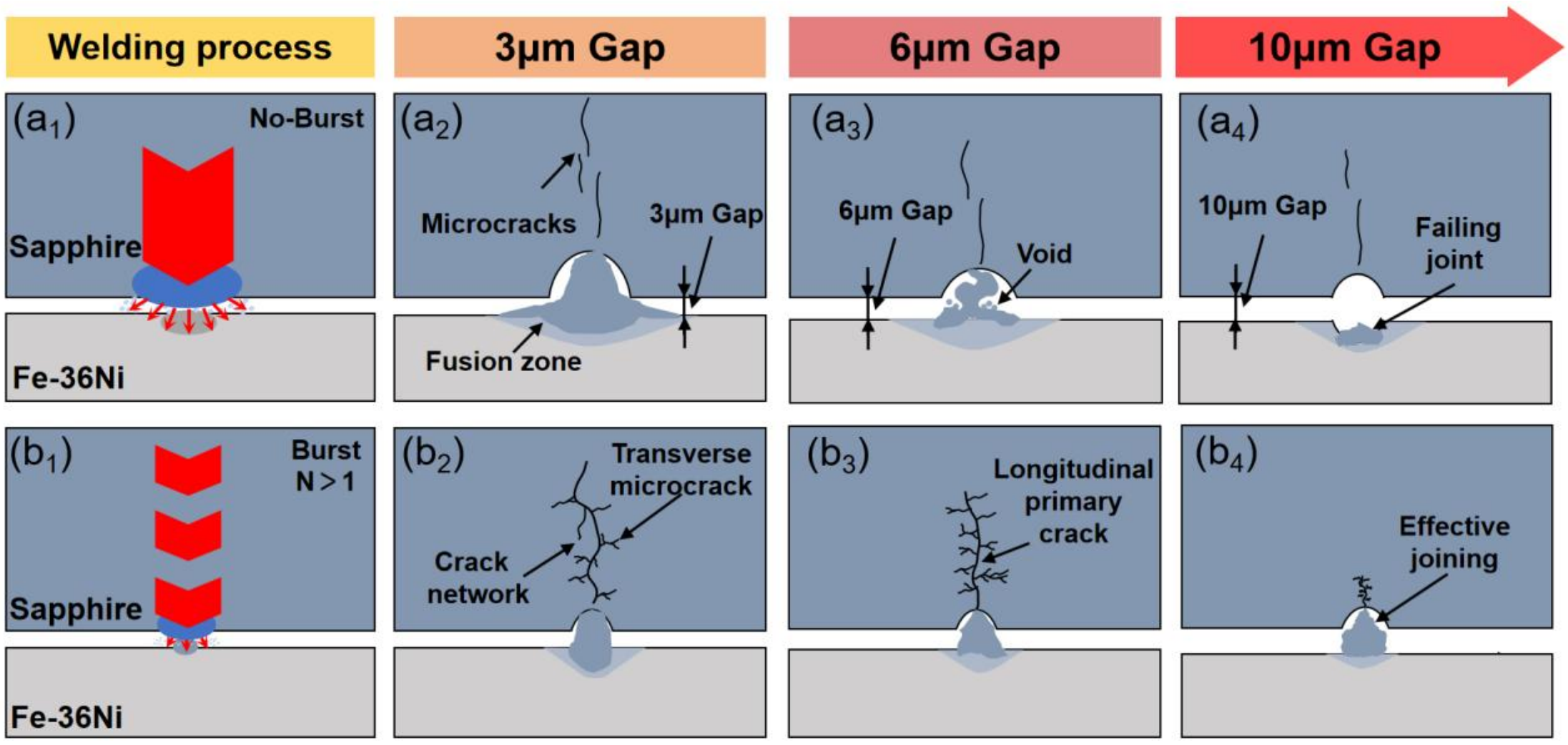

**Fig 13. Schematic illustrations of the welding process and resulting joint microstructure under no-Burst and Burst modes at different gap conditions.** **$a_1$** no-Burst welding process. **$a_2$-$a_4$** joint microstructure after no-Burst welding at varying gaps. **$b_1$** Burst-mode welding process. **$b_2$-$b_4$** joint microstructure after Burst-mode welding at different gaps.

The first mechanism is burst-induced crack damage in sapphire. This damage originates from the intrinsic brittleness of sapphire under cyclic burst-mode thermal loading. Each sub-pulse induces a localized heating–cooling cycle, producing repeated thermal expansion and contraction within the laser-modified region. At small gaps, where interfacial energy coupling is already sufficient, the additional thermal accumulation contributes only marginally to bonding but continuously increases thermal stress in the near-interface sapphire region. Once microcracks are initiated, subsequent sub-pulses can repeatedly reload crack tips, promoting crack extension, branching, and interconnection. Consequently, isolated defects evolve into a dense subsurface crack network.

The second mechanism is burst-enhanced interfacial bridging. This mechanism originates from the temporally distributed energy deposition of burst-mode welding. In the no-burst mode, energy is delivered within a single ultrafast event and is rapidly attenuated by plasma shielding and propagation loss across the air gap. In contrast, burst-mode welding delivers the incident energy through a sequence of temporally separated sub-pulses. The first sub-pulse initiates local melting and plasma formation, whereas subsequent sub-pulses interact with a preheated and partially relaxed interaction zone. This stepwise energy-coupling process reduces transient shielding, promotes cumulative thermal accumulation, prolongs molten-pool lifetime, stabilizes interfacial flow, and sustains elemental interdiffusion. Consequently, burst-mode welding compensates for gap-induced energy loss and enables metallurgical bridging across non-optical-contact gaps, thereby improving gap tolerance.

The competition between these two mechanisms defines three distinct welding regimes.

At a 3 μm gap, the interface is close to the joining limit of conventional no-burst welding, and sufficient interfacial reaction can still be achieved without burst-mode assistance. In this regime, the bridging contribution of burst-induced thermal accumulation is limited because the interfacial reaction threshold has already been reached. Consequently, burst-induced crack damage in sapphire becomes the dominant factor controlling joint failure. Although burst-mode welding improves interfacial continuity and promotes elemental mixing, these benefits are outweighed by the formation of a dense crack network in the brittle sapphire substrate. The failure mode therefore shifts from localized interfacial fracture to crack-network-assisted delamination, resulting in a monotonic decrease in joint strength with increasing number of sub-pulses.

At a 6 μm gap, the two mechanisms became directly competitive, and the process entered a transition regime. In the no-burst mode, single-pulse energy delivery became insufficient after propagation across the widened air gap, leading to unstable melting, incomplete interfacial reaction, and discontinuous bonding with pores and lack-of-fusion defects. Under these conditions, the interfacial-bridging contribution of burst-induced thermal accumulation became comparable to, and began to exceed, its crack-inducing effect. By prolonging molten-pool lifetime and sustaining interfacial heat input, burst-mode welding restored reaction continuity, enhanced elemental interdiffusion, and suppressed gap-induced discontinuities. This marked a critical inversion point where the dominant role of burst-mode welding shifted from damage accumulation to bonding enhancement, allowing the joint strength to exceed that of the no-burst mode.

At a 10 μm gap, the process entered a gap-dominated regime in which conventional no-burst welding failed to form an effective joint. Severe plasma shielding, beam divergence across the air gap, and transmission loss reduced the single-pulse energy below the threshold required for stable interfacial melting. In contrast, the interfacial-bridging effect of burst-mode welding became dominant. Through temporally staged energy deposition, repeated sub-pulse reheating, and sustained molten-pool evolution, burst-mode welding overcame the energy-coupling barrier imposed by the large air gap and progressively established a stable reaction zone across the interface. This enabled continuous gap filling, effective metallurgical bridging, and reliable joint formation across a 10 μm non-optical-contact gap. In this regime, burst-mode welding no longer functioned merely as a thermal modulation strategy but became the key enabling mechanism that transformed the interface from non-joinable to joinable.

Therefore, the essence of large-gap burst-mode welding lies in a gap-dependent transition of the dominant mechanism: from crack-damage-controlled failure at small gaps, to a competitive transition at intermediate gaps, and finally to gap-tolerance-dominated metallurgical bridging at large gaps. This mechanism explains both the strength inversion at intermediate gaps and the extension of the joining limit achieved by burst-mode ultrafast welding under large-gap non-optical-contact conditions.

## 3. Conclusion

This study establishes a mechanism-oriented framework for burst-mode ultrafast laser welding of sapphire/Invar alloy under non-optical-contact conditions, revealing a gap-dependent transition in interfacial response and defining a unified process boundary for large-gap heterogeneous joining.

(1) A gap-dependent competing mechanism is identified as the governing principle of burst-mode welding, where the interfacial response transitions from crack-dominated degradation at small gaps to thermally accumulated bonding enhancement at larger gaps, rather than exhibiting monotonic improvement.

(2) Stable interfacial bonding is successfully achieved under a large 10 μm non-optical-contact gap, exceeding the effective joining limit of conventional no-burst processing. Under the optimized 3-burst condition at 130 μJ, burst-mode welding enables continuous gap bridging through temporal thermal accumulation and achieves a maximum shear strength of 6.3 MPa.

(3) The distinct roles of burst-mode welding are further clarified across different gap scales. At 3 μm, cyclic thermal loading induces localized stress concentration and crack network formation in sapphire. At larger gaps, prolonged molten-pool lifetime promotes interfacial diffusion and metallurgical bridging, leading to a transition from damage-dominated to bonding-dominated behavior.

(4) Multi-scale characterization confirms the proposed mechanism, revealing a consistent structure–property correlation through fracture morphology, elemental redistribution, and interfacial topology evolution.

Overall, this work establishes a gap-controlled design rule for burst-mode laser joining, providing a generalized strategy for achieving robust, gap-tolerant heterogeneous integration beyond conventional joining limits.

# 4.Materials and Methods

## 4.1Materials and Experimental setup

For this experiment, double-side polished sapphire crystals (10 mm × 10 mm × 3 mm, surface roughness Sa = 0.28 nm) and single-side polished Invar alloy (Invar 36, dimensions: 10 mm × 10 mm × 2 mm, surface roughness Sa = 27.52 nm) were used as the welding materials. The key thermophysical property parameters of the two materials are listed in Table 1. The laser source was a tunable repetition-rate 1064 nm picosecond pulse laser (Amber NX IR-50) with a pulse width of 15 ps and a maximum repetition frequency of up to 1 MHz. In the experiments, no-burst mode and burst-mode were employed for the welding process. The specific process parameters are listed in Table 2. The laser scanning followed a linear path to form the weld seam. Its cross-sectional morphology was obtained through linear scanning, while the weld shape for shear testing was rectangular as shown in Fig. 14d.

**Table 1. Thermal and physical properties of Invar 36 and sapphire**

| Material | Melting point (°C) | Coefficient of thermal expansion (10−6 /K) | Tensile or breaking strength (MPa) | Modulus of elasticity (GPa) |
|---|---|---|---|---|
| Sapphire ($Al_2O_3$) | 2040 | 5.0-6.7 | 400 | 345 |
| Invar 36(Fe-36Ni) | 1427 | 1.2-2.0 | 488 | 141 |

**Table 2. Laser parameters of Ultrafast laser welding sapphire and Fe-36Ni alloy**

| Laser parameter | Value |
|---|---|
| Central wavelength (nm) | 1064 |
| Pulse width (ps) | 15 |

| Repetition rate (KHz) | 251 |
| --- | --- |
| Pulse energy (μJ) | 55/70/85/100/115/130 |
| Scanning speed (mm/s) | 23 |
| Number of sub-pulses | 1/3/5/7 |
| gap size(μm) | 3/6/10 |

**Fig 14. Schematic diagrams of the experimental setup, laser modes, and testing method. a,** Picosecond laser welding of sapphire ($\alpha$-$Al_2O_3$) and Invar (Fe-36Ni alloy)**. b,** Sample clamping configuration. **c,** Shear strength test fixture. **d,** Laser scanning paths. **e,** no-Burst (single-pulse) mode. **f,** Burst mode principle. **g,** Temporal profile of a 3-Burst mode.

After welding, the samples were sectioned using wire cutting, followed by grinding and polishing to prepare cross-sectional specimens for microstructural characterization. A 3D digital microscope (VHX600, Keyence) was used for surface morphology observation. Scanning electron microscopy (SEM, SEM5000, CIQTEK Co., Ltd.) equipped with energy-dispersive X-ray spectroscopy (EDS, Ultim Max 40, Oxford Instruments) was employed for interface morphology and elemental analysis of the welded joints.

Shear strength tests were conducted using a universal testing machine (UTM5105, Shenzhen SUNS) equipped with a custom fixture, as shown in Fig 14c, applying a loading rate of 0.2 mm/min. The joint strength was calculated using the formula $\sigma = F / S$, where F is the maximum shear load and S is the actual bonding area. After fracture, a scanning white-light interferometer was used to reconstruct the 3D topography of the fracture surfaces for quantitative morphology analysis.[44]

To accurately control the welding process under different gap conditions resulting from varying assembly accuracies, this study introduced polyimide films with specific thicknesses in the range of 3-10 μm as spacers, as shown in Fig. 15. During the experiment, two rectangular polyimide films were placed at the left and right ends between the two samples to be joined. A dedicated fixture was used to clamp the transparent and metal materials together, thereby precisely controlling and maintaining the interfacial gap in the welding region at the nominal thickness of the spacer. The interfacial gap was then measured and verified using a white-light interferometer.[25] Subsequently, picosecond laser welding was performed, followed by multi-scale characterization and destructive shear testing.

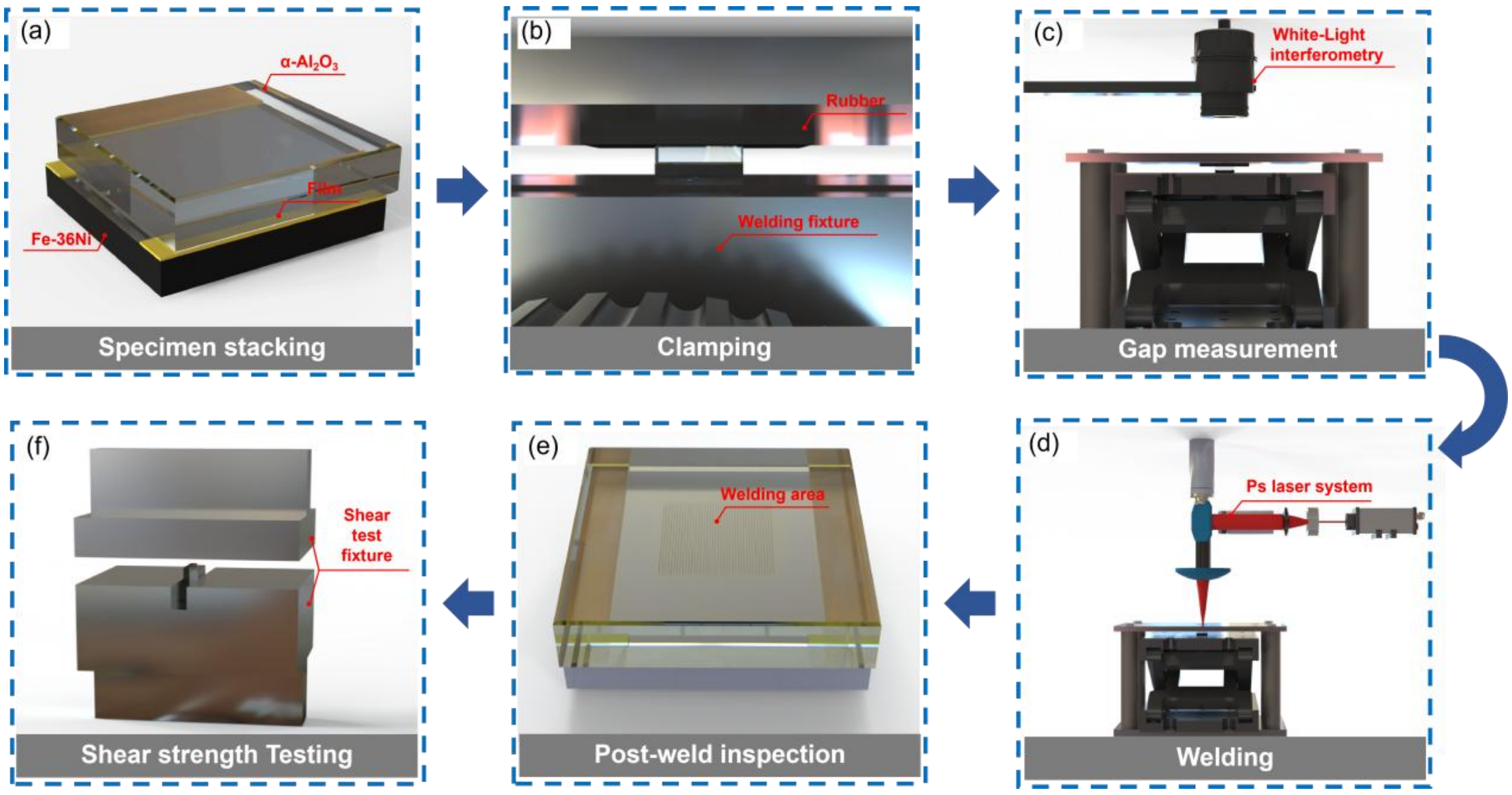

**Fig 15. Schematic of the experimental procedure. a,** Sequential stacking of Invar ally, spacer film, and sapphire. **b,** Clamping the assembly with a dedicated fixture. **c,** Measuring the interfacial gap via white-light interferometry. **d,** Welding with a picosecond (ps) laser. **e,** Inspecting the weld seam. **f,** Shear strength testing.

## Acknowledgements

This research was supported by the National Key Research and Development Project of China (2023YFB4605503). The Open Project Program of Key Laboratory for Cross-Scale Micro and Nano Manufacturing, Minstry of Education, Changchun

University of Science and Technology (CMNM-KF202403).

**Author Contributions**

S.Y. conceived the idea and supervised the project. Y.L. and N.L. constructed the experimental system and performed the welding experiments. Y.L., N.L., and Y.C. contributed to sample preparation, multi-scale characterization (including SEM, EDS, and white-light interferometry), and data analysis. Y.W. developed the method for simulating continuous gradient gap variation. All authors participated in data analysis and contributed to writing the manuscript.

**Data availability**

All data are available from the corresponding authors upon reasonable request.

**Conflict of interest**

The authors declare no competing interests.